\documentclass{article}
\usepackage[utf8]{inputenc}
\usepackage{amsmath}
\usepackage{amsfonts}
\usepackage[T1]{fontenc}
\usepackage[toc,page]{appendix}
\usepackage{XCharter}
\usepackage[xcharter,bigdelims,vvarbb]{newtxmath}
\usepackage{booktabs}
\usepackage[noblocks,affil-sl]{authblk}
\usepackage{siunitx}
\usepackage{listings}
\usepackage[table]{xcolor}
\usepackage[colorlinks = true,
            linkcolor = blue,
            urlcolor  = blue,
            citecolor = blue,
            anchorcolor = blue]{hyperref}
\usepackage{caption}
\usepackage{subcaption}

\author{Bálint Máté}
\affil{Department of Computer Science, University of Geneva, 1227 Carouge, Switzerland}

\author{Bertrand Le Saux}
\affil{European Space Agency (ESA) $\Phi$-lab, 00044 Frascati (RM), Italy}

\author{Maxwell Henderson}
\affil{Independent Researcher}

\definecolor{codegreen}{rgb}{0,0.6,0}
\definecolor{codegray}{rgb}{0.5,0.5,0.5}
\definecolor{codepurple}{rgb}{0.58,0,0.82}
\definecolor{backcolour}{rgb}{0.95,0.95,0.92}

\lstdefinestyle{mystyle}{
    backgroundcolor=\color{backcolour},   
    commentstyle=\color{codegreen},
    keywordstyle=\color{magenta},
    numberstyle=\tiny\color{codegray},
    stringstyle=\color{codepurple},
    basicstyle=\ttfamily\footnotesize,
    breakatwhitespace=false,         
    breaklines=true,                 
    captionpos=b,                    
    keepspaces=true,                 
    numbers=left,                    
    numbersep=5pt,                  
    showspaces=false,                
    showstringspaces=false,
    showtabs=false,                  
    tabsize=2
}

\title{\textit{Beyond Ansätze: Learning Quantum Circuits \\as Unitary Operators}}
\date{\today}
\usepackage{algorithm}
\usepackage[margin=1in]{geometry}
\usepackage{graphicx}
\usepackage[noend]{algpseudocode}
\usepackage[square]{natbib}
\usepackage[T1]{fontenc}
\usepackage[utf8]{inputenc}

\newtheorem{theorem}{Theorem}
\newtheorem{question}[theorem]{Question}
\begin{document}

\maketitle

{{\let\thefootnote\relax\footnotetext{Code for the experiments is available at \url{https://github.com/balintmate/beyond-ansaetze}}}

 {{\let\thefootnote\relax\footnotetext{Corresponding author: balint.mate@unige.ch}}

\begin{abstract}
    This paper explores the advantages of optimizing quantum circuits on $N$ wires as operators in the unitary group $U(2^N)$. We run gradient-based optimization in the Lie algebra $\mathfrak u(2^N)$  and use the exponential map to parametrize unitary matrices. We argue that $U(2^N)$ is not only more general than the search space induced by an ansatz, but in ways easier to work with on classical computers. The resulting approach is quick, ansatz-free and provides an upper bound on performance over all ansätze on $N$ wires.
\end{abstract}
\section{Introduction}

Quantum machine learning (QML) continues to be one of the most compelling application areas of quantum computing, particularly in the noisy, intermediate scale quantum (NISQ) era \citep{preskill_nisq}. The field has already seen a broad range of QML applications investigated, including image classification \citep{quantum_kitchen_sinks, Adachi2015}, predicting quantum states associated with a one-dimensional symmetry-protected topological phase \citep{Cong2019}, election forecasting \citep{Henderson2019}, financial applications \citep{Alcazar2019, Kashefi2020}, synthetic weather modeling \citep{Enos2021} or Earth observation~\citep{qnn, sebastianelli-QNN4EO-JSTARS2021}. The unique properties of quantum computers powering QML applications are tested against classical algorithms, with the goal of observing higher accuracy, faster training, fewer required training samples, or other beneficial improvements.

While some QML applications have theoretical advantages over classical algorithms, empirically showing advantages on NISQ devices remains a challenge for several reasons. One issue is in optimal approaches for embedding classical data into QML devices, since noise can lead to incorrect, unstable outputs. Research into methods for determining which embeddings minimize the impact of this noise for particular problems offers an appealing approach forward \citep{LaRose2020}. Another problem is that of data dimensionality, as NISQ devices are extremely limited in qubit count compared to the typical dimensionality of input classical data. Various approaches to mitigate this limitation have been explored, including linear encoding a large amount of classical data into a few control parameters \citep{quantum_kitchen_sinks}, running the circuit on subsets of the entire input \citep{qnn, Henderson2021}, and re-uploading data using deeper circuits \citep{Perez-Salinas2019}. While there are additional obstacles in performing QML applications in practice, we finish with an extremely important one and the focus of this paper: dealing with circuit ansatz selection and training times.

The circuit ansatz is the structure of the quantum gates that comprise the quantum circuit, which fundamentally define the types of functions that the quantum circuit can compute; this is analogous to the types of layers that form the architecture of a neural network. The ansatz can be considered as an extra assumption on the form of the problem and usually comes from an educated guess. Finding good circuit ansätze which are effective on near term devices is extremely challenging for QML applications, as it is often unclear for various datasets how different circuit ansätze will perform. Additionally, once a circuit ansatz is chosen, it will typically have tunable parameters which need to be optimized in order to generate the best performance possible. Methods of parameter-shift rule have been developed and improved upon for calculating quantum gradients to update these parameters \citep{quantum_circuit_learning, Schuld2018}, but the computational cost becomes extremely taxing even using state-of-the-art software implementations \citep{bergholm2020pennylane} as the problem size scales up. Ideally, better approaches for determining optimal circuit ansatz with hardware amenable structure and low numbers of variational parameters would be ideal.

This work focuses on a methodological approach which uses powerful machine learning libaries to try and learn the coefficients of unitary matrices. The goal is to provide an approach which can theoretically hone in on extremely good, problem-specific unitary mappings, that allow researchers to qucikly estimate the performance of QML applications without having to choose an ansatz. In Section \ref{sec:motivation}, we walk through a more detailed account of current problems and solutions for selecting and training circuit ansatz for QML applications. In Section \ref{sec:method}, we explain the approach laid out in this work, and show the performance of our approach compared to other best-in-class software in Section \ref{sec:experiments}. Finally, in Section \ref{sec:future_work_conclusion} we address ways in which this work may be extended and improved upon that could further benefit the QML research community.

\section{Motivation} \label{sec:motivation}
\subsection{Existing quantum simulations are slow}

While the future of quantum computation is in quantum computational hardware, the currently available hardware is expensive and accessibly limited; this is noted at the core reason for the work of \citet{Padilha2019}, which is one of many advances in designing classical software simulators of quantum computational systems. While these quantum simulators allow the research community to experiment on toy problems, even the best simulators pose major problems for many QML applications.

Focusing in on the circuit ansatz challenge, we may consider on the application area of quantum chemistry. Within quantum chemistry, there are some some algorithms wherein ``problem-inspired" circuit ansatz exists, but due to the current quantum computational hardware constraints, a ``hardware efficient ansatz" is used instead. This is a practical compromise; a circuit ansatz which performs well in simulation but fits no current hardware may be traded off for a circuit ansatz which has a slew of theoretical problems but can be experimented with on NISQ devices.

QML applications in this sense have an extreme disadvantage: unlike quantum chemistry problems which are built on strong foundations, there are no known theoretical ansatz which work well on arbitrary machine learning problems. By their very nature, machine learning applications can have arbitrary datasets with patterns that do not match any clear physical model. In this case,  there is no clear answer between ``what does the ideal anstaz look like?" vs ``what is close to the ideal we can fit on hardware?" but rather ``what ansatz is useful on this particular dataset within this particular application?"

\subsection{A quantum circuit is just a unitary transformation}
The key observation that motivated this paper is that a quantum system of $N$ qubits can be described by a state vector in $\mathbb{C}^{2^N}$ and every quantum circuit is a unitary transformation on this state space. This implies that every machine learning algorithm that optimizes a quantum circuit on $N$ wires can be viewed as a optimization process in (a subspace of) the space of $2^N$-dimensional unitaries, denoted by  $U(2^N)$. Note that this a fundamentally different from classical machine learning where the space of all possible networks is infinite dimensonal. In the quantum case, the unitary group $U(2^N)$ has real dimension $2^{2N}$ and allows to fully parametrize the possible quantum circuits on $N$ wires.

\section{Unitary optimization} \label{sec:method}
We want to perform gradient descent in the space of unitary matrices without having to choose an ansatz for the architecture. 
Such an approach would have the advantage of not  losing any expressivity of the quantum network by constraining the optimization to a fixed architecture.

Moreover, directly optimizing the unitary enables the full use of differentiable programming frameworks. In particular, we can efficiently process batch of inputs in a single pass of the neural network and make use of GPUs for faster training.

\subsection{Parametrizing via the Lie algebra $\mathfrak {u}(d)$}
The Lie algebra $\mathfrak {u}(d)$ of the unitary group $U(d)$ is the space of $d \times d$ skew-Hermitian matrices
$$\mathfrak u(d)=\{X \in M_d(\mathbb{C})|X=-X^\dagger\}$$
This space is of real dimension $d^2$.
Moreover, since $U(d)$ is connected and compact, the (matrix-)exponential map $\mathfrak {u}(d) \rightarrow U(d)$ is surjective \citep[Corollary 11.10]{hall2015lie}, i.e. every unitary transformation $A \in U(d)$ can be written as the exponential of some skew-Hermitian $X \in \mathfrak u(d)$.

Motivated by this result, we propose to parametrise the Lie algebra $\mathfrak u(d)$, and use the exponential map to obtain unitary transformations. Since the exponential map is differentiable, we can propagate gradients all the way back to the Lie algebra and do gradient based optimization there. 

Note that is significantly eases the optimization process. The reason for this is that the Lie group of unitary transformations has non-trivial geometry and doing gradient descent is cumbersome on such spaces. On the other hand, the Lie algebra is a vector space, where gradient descent can shine.

Since PyTorch \citep{paszke2017automatic} has built-in support both for complex-valued tensors and the matrix exponential, random unitary transformations can be generated with just a few lines of Python code:
\lstset{language=Python}
\lstset{frame=lines}
\lstset{caption={Generating a random unitary transformation in PyTorch}}
\begin{lstlisting}
N  = 5                                       # number of qubits
D  = 2**N                                    # dimensionality of the quantum system
x  = torch.randn(D,D, dtype=torch.cfloat)    # random (D,D) complex matrix
sH = x - x.T.conj()                          # sH is skew-hermitian
U  = torch.matrix_exp(sH)                    # U  is unitary
\end{lstlisting}
Parametrising a unitary transformation on $D$ dimensions requires $D^2$ scalar parameters, and the dimension of an $N$ qubit system is $D=2^N$. It follows that the use our approach on $N$ wires requires $2^{2N}$ trainable parameters.
\subsection{Reducing the parameter count}
\label{sec:subsets}
Since the dimension of the group of unitary transformations scales exponentially as we increase the number of qubits, it is impractical to optimize unitary operators on a large number of wires. Restraining to optimization to subsets of $U(2^N)$ can ease this computational issue.
One interesting subgroup is $U(2^{k}) \otimes U(2^{N-k})$, i.e. applying one unitary to the first $k$ wires, and a second one to the last $N-k$ wires. This of course can be generalized by a different partitioning of the $N$ wires.
For instance, consider the following alternatives:
\begin{itemize}
    \item Optimizing directly on $N$ qubits requires $2^{2N}$ parameters to train.
    \item Optimizing $N/k$ unitary operators on $k$ wires requires $(N/k)2^{2k}$ parameters. This, of course, does not allow any interaction between qubits belonging to different partitions. To overcome this, one could repeat the same procedure with a different partitioning of the wires into groups of size $k$. Repeating this $m$ times results in $(Nm/k)\,2^{2k}$ trainable parameters, which only scales linearly in $N$.
\end{itemize}
Working with subspaces of the unitary group in order to shrink the optimization space is an effective way of cutting the computational costs. It is important to note that every such reduction of the number of parameters (and in turn the search space) trades off generality for a computational speedup. Moreover, training speed is not a monotonically decreasing function of the parameter count, a large number of parameters may be faster if the resulting map is simple to compute (such as the exponential of a single skew-symmetric matrix) than a mapping with a lower parameter count but with a long chain of simple ingredients, that need to be composed every time the parameters change (such as a long ansatz).

\subsection{Connection to circuits with a given ansatz}
Taking the idea of the previous section to the extreme, we can recover the traditional approach of first choosing an ansatz of predefined quantum gates and optimizing their parameters. 
As an example, consider a two-qubit quantum system. Our approach would optimize the parameters $(\theta_1,\, ...\,, \theta_{16}) \in \mathbb{R}^{16}$ generating the full unitary group $U(4)$:
\begin{equation}
    \label{eq:full4x4}
    \Bigg\{\exp
\begin{bmatrix}
  \theta_1 i            & \theta_2+\theta_3i       & \theta_4+\theta_5i       & \theta_6+\theta_7i\\ 
  -\theta_2+\theta_3i   & \theta_8 i               & \theta_9+\theta_{10}i     & \theta_{11}+\theta_{12}i\\
  -\theta_4+\theta_5i   &-\theta_9+\theta_{10}i    & \theta_{13}i               & \theta_{14}+\theta_{15}i\\ 
  -\theta_6+\theta_7i   &-\theta_{11}+\theta_{12}i & -\theta_{14}+\theta_{15}i & \theta_{16}i\\ 
\end{bmatrix} \Bigg|(\theta_1,\, ...\,, \theta_{16}) \in \mathbb{R}^{16}
\Bigg\}
\end{equation}
where $i$ denotes the complex unit.

Now, let us restrict our attention to the ansatz with an RX-gate on the first wire and an RY-gate on the second one. Optimizing this circuit amounts to training $(\theta_1,\theta_2) \in \mathbb{R}^2$ resulting in the following family of unitaries:
\begin{equation}
    \label{eq:ansatz4x4}
\Big\{\begin{pmatrix}
  \cos \theta_1/2 & -i\sin \theta_1/2\\ 
  -i\sin \theta_1/2 & \cos \theta_1/2
\end{pmatrix} \otimes 
\begin{pmatrix}
  \cos \theta_2/2 & -\sin \theta_2/2\\ 
  \sin \theta_2/2 & \cos \theta_2/2
\end{pmatrix}\Big|(\theta_1,\theta_2) \in \mathbb{R}^2 \Big\}
\end{equation}
Note that family of unitaries in (\ref{eq:ansatz4x4}) is a strict subset of the unitaries in (\ref{eq:full4x4}), i.e. our method can find quantum circuits that the approach with a given ansatz would not be able to find. The price to pay for that is an increased number of parameters, a bigger search space to optimize in. 

As with all ansätze, if there is a good reason to believe that the optimal circuit belongs to a certain subclass of unitaries, then it makes a lot of sense to to only consider functions that satisfy the ansatz. Otherwise, keeping the search space as big as possible (and practically still feasible) is a reasonable thing to do. 

\subsection{Bound on the performance of all ansätze}
\label{sec:bound}
In the previous section we argued that an ansatz is equivalent to a particular restriction of the search space. Conversely, optimization over all unitaries is at least as general as any choice of ansatz, which in turn implies that the optimum of full unitary optimization represents the best possible quantum circuit that any ansatz could have found\footnote{Note that in practice life is more complicated. Local minima, gradient step size, etc. can have an effect on the optimization process. The statement here is simply that if a search space is contained in an other one, then the \textit{global optimum} of the latter one cannot be worse than that of the first one.}.
To put it differently:
\begin{theorem}
Optimization in the full unitary group $U(2^{2N})$ provides an upper bound on performance over all possible ansätze on $N$ wires.
\end{theorem}

This is particularly useful when trying to answer questions such as 
\begin{question}
\label{Q}
Given some machine learning task $\mathcal{T}$, is there a quantum circuit on $N$ wires that performs better than some predefined  performance threshold $\mathcal{P}$?
\end{question} 
Whether  $\mathcal{P}$ is the performance of some classical approach (looking for quantum advantage) or not, our method separates these issues from the choice of an ansatz and provides a way to answer them easily. If the answer to a question such as Question \ref{Q} is affirmative, there is hope that a (simple) ansatz can also perform well on $\mathcal{T}$. Otherwise one doesn't have to worry about ansätze as none of them will reach performance $\mathcal{P}$.

\section{Experiments} \label{sec:experiments}
\subsection{Toy experiments} \label{exp:1}
We compare our approach to PennyLane \citep{bergholm2020pennylane} by running the same training procedure (number of qubits, datapoints and training epochs) in both frameworks. 
The space of unitary transformations on $N$ qubits is $2^{2N}$ dimensional. It follows that our PyTorch model contains $2^{2N}$ trainable parameters. 
Accordingly, we initizalize a  PennyLane model containing a single RandomLayer with $2^{2N}$ trainable parameters.

In our experiments, we generate a classical dataset and use $R_x$ rotations to encode the classical data into a quantum representation, apply the quantum circuit and perform $Z-$measurements on the wires to decode from the quantum to a classical representation. We are not interested in accuracy of the trained networks, only in training speed, so we just train the networks to learn the identity function.
\subsubsection*{Comparing to PennyLane}
In the first experiment, the dataset is just 1 datapoint of some fixed dimension $N$.  We train for 10 epochs both in PennyLane and PyTorch and log the wall-clock time after every epoch. The results are plotted in Figure (\ref{fig:PennyLane_PyTorch}). Our approach is significantly faster than PennyLane which we explain by the fact that our approach optimizes a single unitary matrix and does not have to compose exponentially many gates.

\begin{figure}[H]
     \centering
     \begin{subfigure}[b]{0.45\textwidth}
         \centering
         \includegraphics[width=\textwidth]{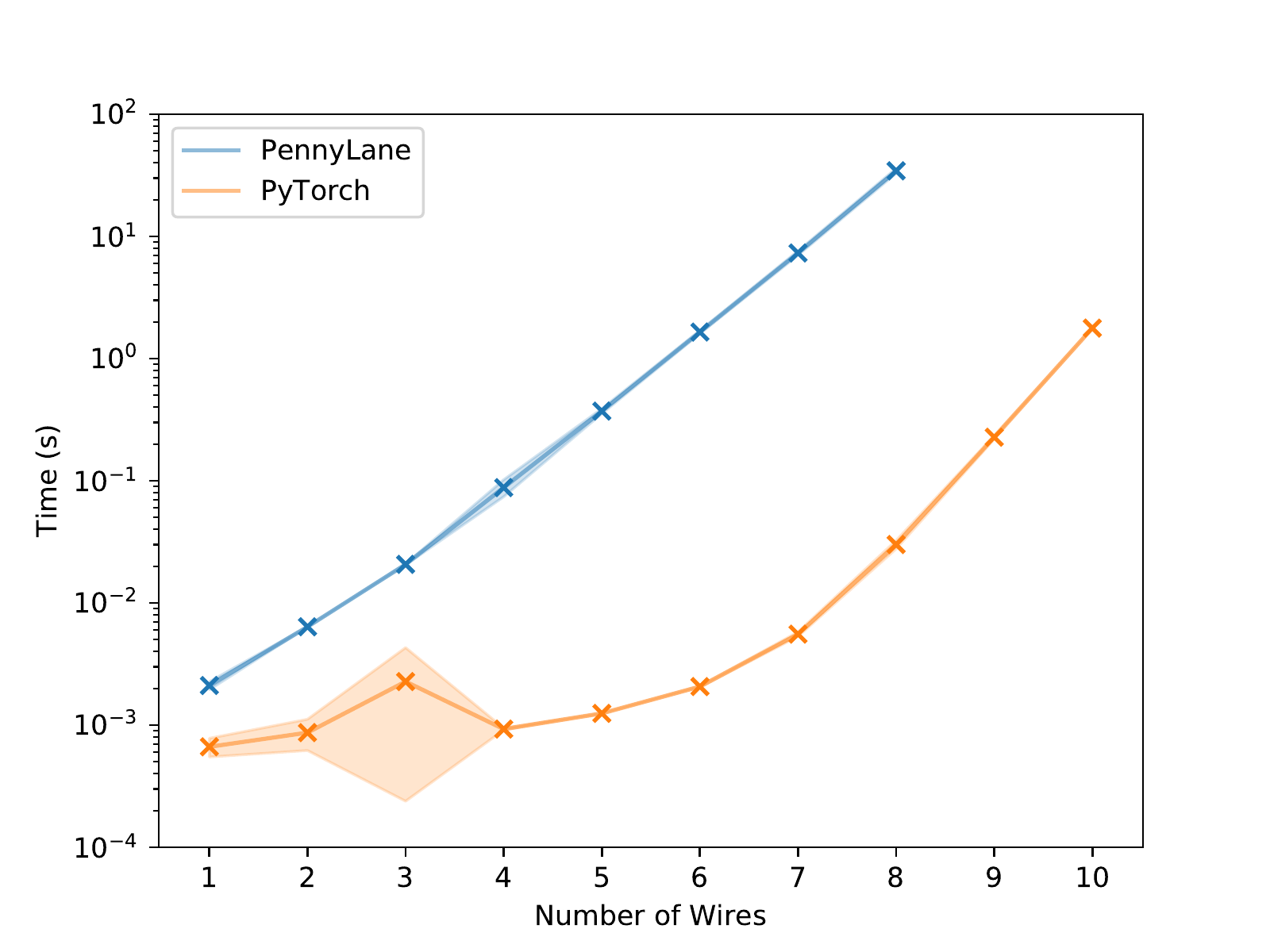}
         \caption{Comparing to PennyLane}
         \label{fig:PennyLane_PyTorch}
     \end{subfigure}
     \hfill
     \begin{subfigure}[b]{0.45\textwidth}
         \centering
         \includegraphics[width=\textwidth]{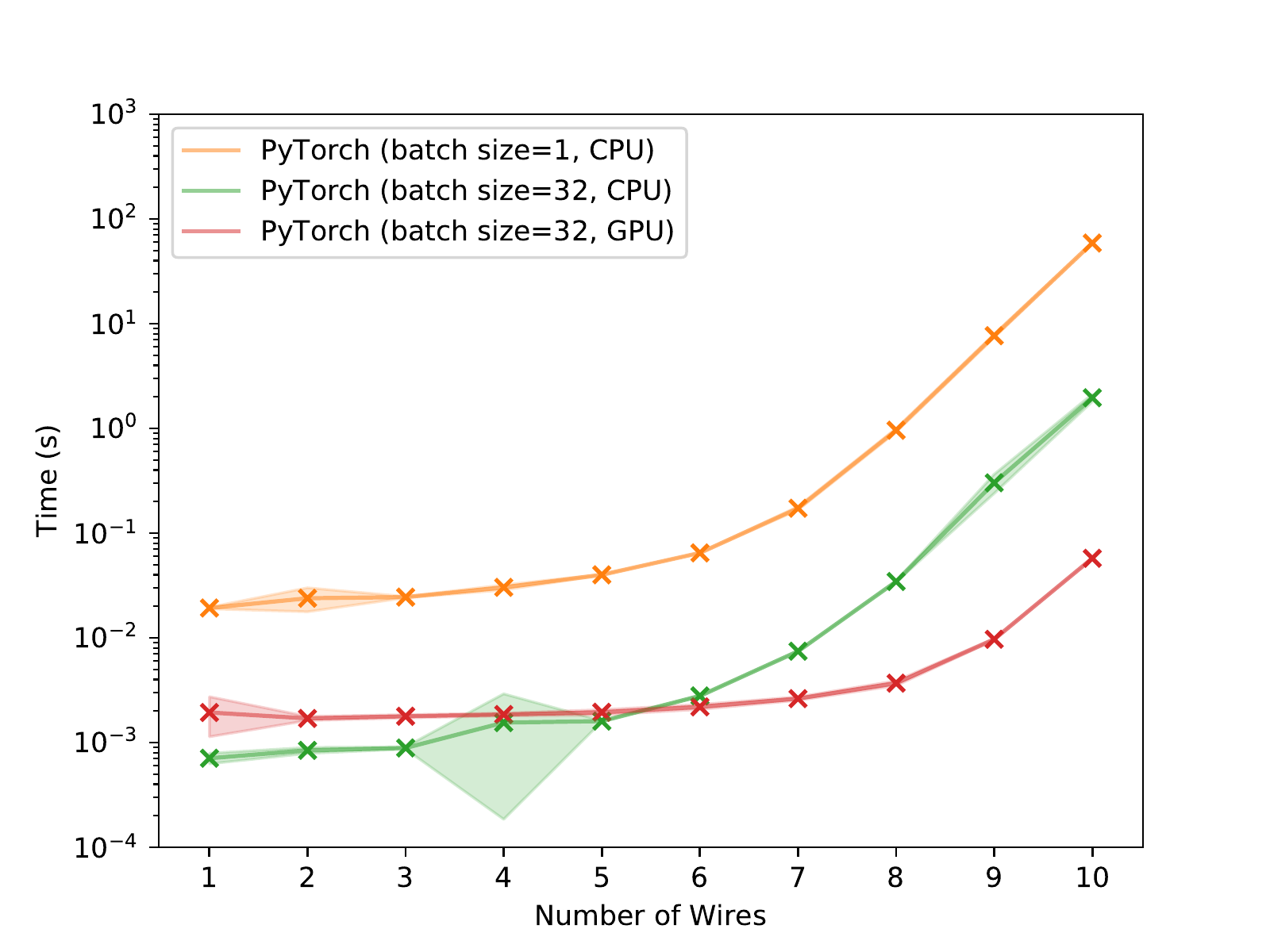}
         \caption{The effect of batching and using GPUs}
         \label{fig:GPU}
    
     \end{subfigure}
    \caption{Results of experiment \ref{exp:1}. On the x-axis the number of qubits vary, while the y-axis shows the average time needed to finish 1 training epoch (average over 10 epochs). Solid lines denote the average, shaded regions the $\pm$ standard deviation. Since the variance bursts are only visible at small circuit size, and are in the order of milliseconds, we interpret them as noise.}     
\end{figure}
\subsubsection*{The effect of batching and using GPUs}
We gain an additional speedup by minibatch training and by using a GPU. To demonstrate this, we build a dataset of 32 datapoints and we train for 10 epochs using 3 different setups:
\begin{itemize}\setlength\itemsep{.5pt}
    \item[1.] PyTorch with batch size 1, train on the CPU
    \item[2.] PyTorch with batch size 32, train on the CPU
    \item[3.] PyTorch with batch size 32, train on the GPU
\end{itemize}
The results are plotted in Figure~(\ref{fig:GPU}). Batching consistently improves speed by a factor comparable to the batch size. Working on the GPU also helps if the circuit is large enough. We explain this by the overhead of invoking GPU kernels, and copying data between CPU and GPU. It's only worth to pay this price above a certain number of wires.
All numerical values are reported in Table~(\ref{table:results}).

\subsection{Quanvolutional networks on MNIST and SAT-6}
\label{exp:2}
In this section we run experiments along the lines of \S\ref{sec:bound} and test quantum circuits on two image classification tasks, MNIST and SAT-6\footnote{The SAT-6 dataset \citep{basu2015deepsat} consists of satellite images of 4-channels (red, green, blue, near infrared)  each belonging to one of 6 categories (barren land, trees, grassland, roads, buildings and water bodies).}.
Quanvolutional networks \citep{qnn} generalize the concept of convolutions by having a quantum circuit sliding through the input image and evaluating it at every patch of the input.
By construction they inherit the translation equivariance property of convolutional networks, therefore are a fitting architecture for image processing tasks.

In both experiments we compare two networks, one classical and one classical-quantum hybrid. The overall number of parameters of the networks are comparable, the architectures are shown in Figure \ref{fig:architecture}. The quanvolutional layer maps from 16 to 8 channels using 2-by-2 filters,  resulting in a total of 32 quantum circuits on 4 qubits. 

The results are summarized in Figure \ref{fig:mnist_sat6}. We see that the quanvolutional architecture trained by full unitary optimization does not yield an accuracy benefit over the classical network and conclude that no ansatz exists for the 4 qubit circuits that would outperform the classical network given that the classical part of the hybrid architecture is not changed, only retrained with the ansatz.

\begin{figure}[H]
     \centering
     \begin{subfigure}[b]{0.48\textwidth}
         \centering
         \includegraphics[width=\textwidth]{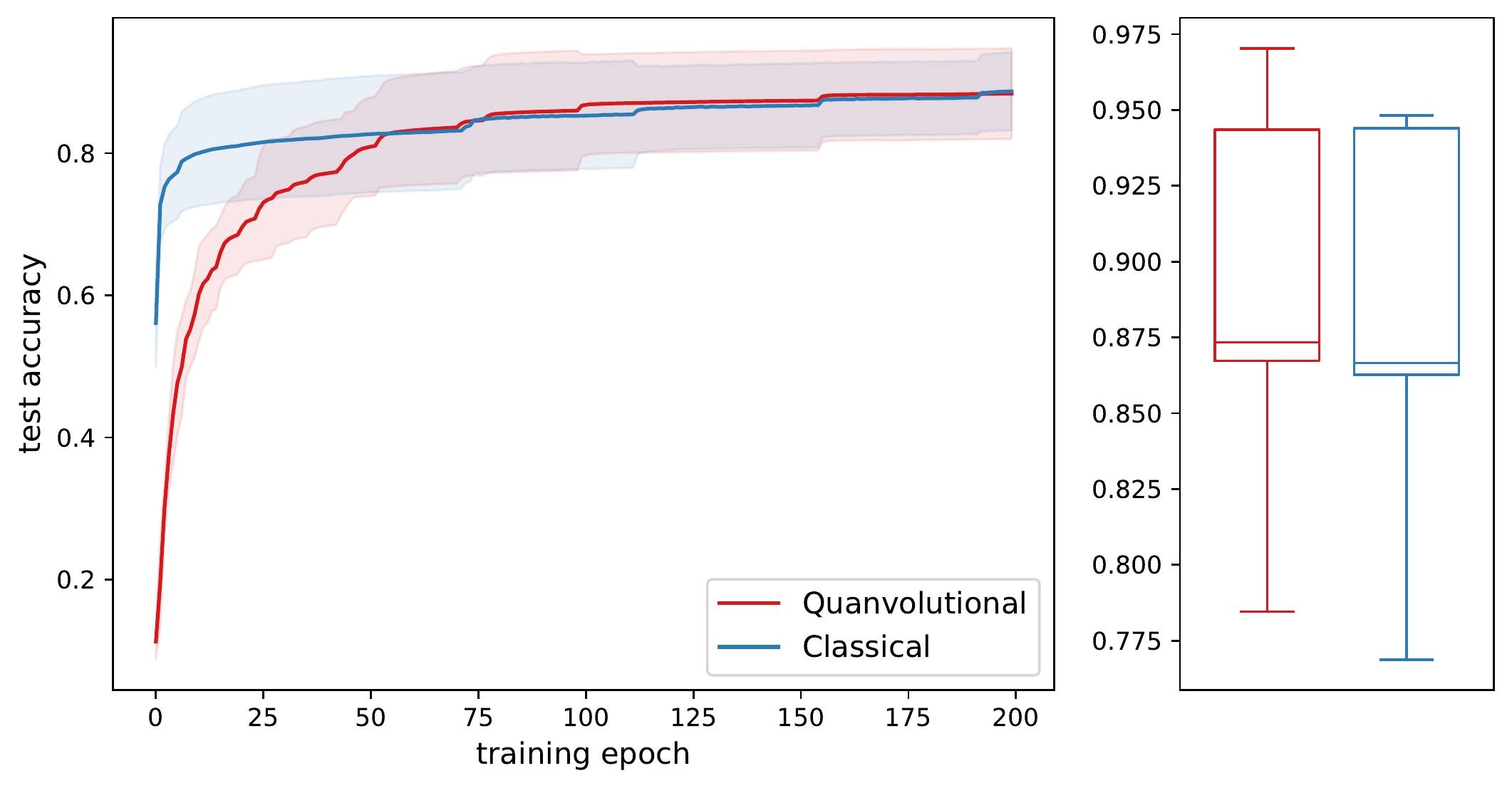}
         \caption{MNIST}
         \label{fig:mnist}
     \end{subfigure}
     \hfill
     \begin{subfigure}[b]{0.48\textwidth}
         \centering
         \includegraphics[width=\textwidth]{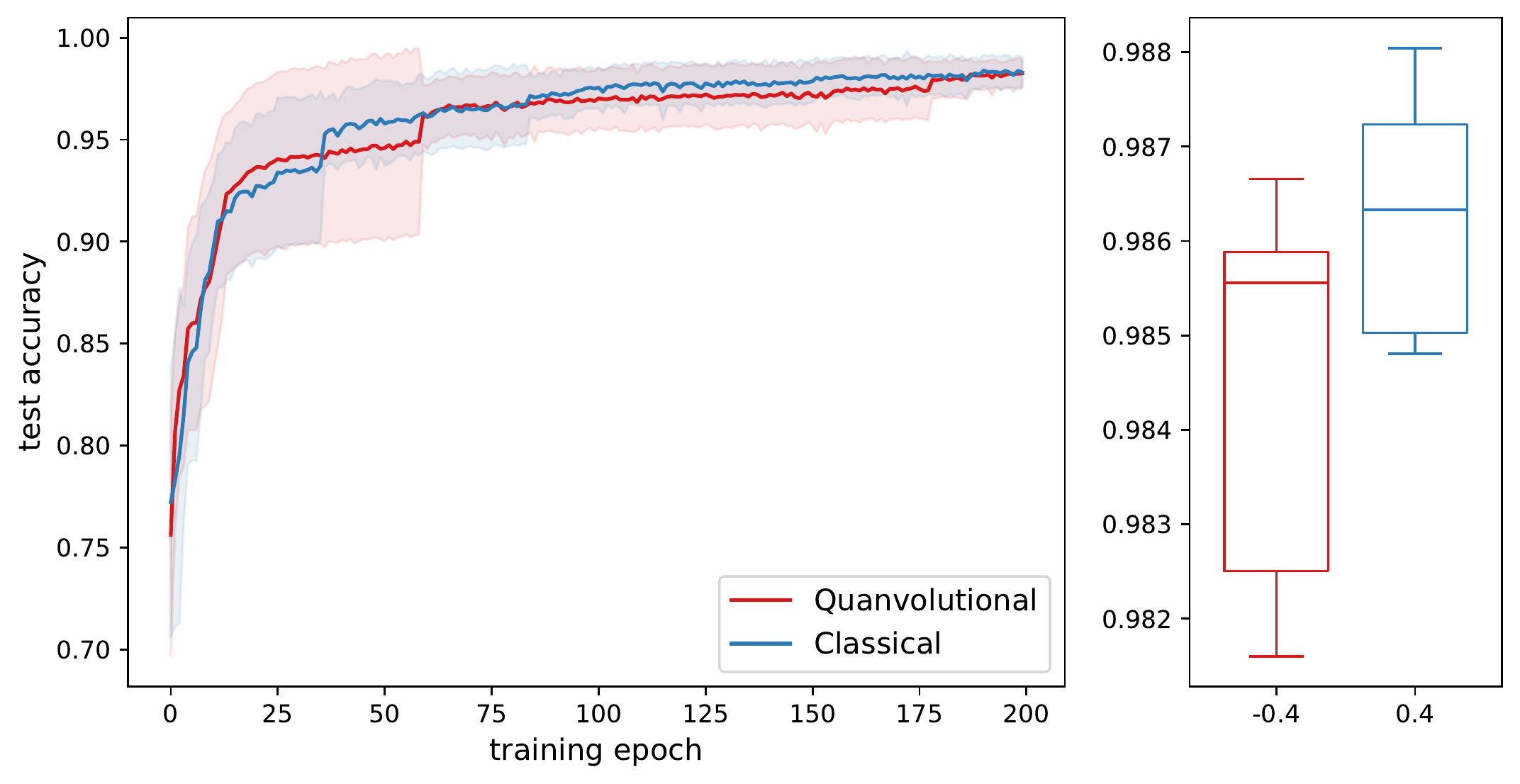}
         \caption{SAT-6}
         \label{fig:sat6}
    
     \end{subfigure}
    \caption{Results of Experiment \ref{exp:2}. The curves display the evolution of the accuracy of the architectures. Solid lines denote the average over 10 runs, shaded regions the $\pm$ standard deviation. The boxplots show the distribution of the accuracy of the architectures after 200 epochs.}
    \label{fig:mnist_sat6}
\end{figure}

\begin{figure}[H]
     \centering
     \begin{subfigure}[b]{0.45\textwidth}
     \centering
            \lstset{language=Python}
            \lstset{frame=lines}
            \lstset{caption={The quanvolutional architecture}}
            \begin{lstlisting}
nn.Conv2d(28,16,4,2),nn.ReLU(),
nn.Conv2d(16,16,3,1),nn.ReLU(),
nn.Conv2d(16,16,2,2),nn.ReLU(),
QConv2d(16,8,2,1),
nn.Flatten(1),
nn.Linear(128,32),nn.ReLU(),
nn.Linear(32,NumC),nn.Softmax(1)\end{lstlisting}
     \end{subfigure}
     \hfill
     \begin{subfigure}[b]{0.45\textwidth}
     \centering
            \lstset{language=Python}
            \lstset{frame=lines}
            \lstset{caption={The classical architecture}}
            \begin{lstlisting}
nn.Conv2d(28,32,4,1),nn.ReLU(),
nn.Conv2d(32,32,3,1),nn.ReLU(),
nn.Conv2d(32,16,2,2),nn.ReLU(),
nn.Conv2d(16,8,2,1),nn.ReLU(),
nn.Flatten(1),
nn.Linear(128,32),nn.ReLU(),
nn.Linear(32,numC),nn.Softmax(1)\end{lstlisting}
    
     \end{subfigure}
    \caption{The two architectures of Experiment \ref{exp:2}. The 4 parameters of \textsl{nn.Conv2d} and \textsl{QConv2d} are \textsl{in\_channels, out\_channels, kernel\_size and stride}, respectively. The variable \textsl{numC} denotes the number of classification categories (10 for MNIST, 6 for SAT-6).}
    \label{fig:architecture}
\end{figure}

\section{Future Work and Conclusion}
\label{sec:future_work_conclusion}

\subsection{Decomposing unitary transformations to canonical quantum gates}
\label{sec:decomposing}
If our approach finds a useful unitary transformation, it needs to be decomposed into a sequence of elementary operations to execute it on a quantum computer. This section summarizes the relevant results.

Any unitary transformation on $N$ wires can be written as a composition of 1 and 2-qubit gates \citep[4.5.2]{nielsen_chuang_2010}. An exact decomposition of a general unitary operator will almost surely not have a simple form, and would require many gates to implement it. However, at the moment we need very shallow circuits to avoid noise challenges. Therefore, a more fitting approach is to do an approximate decomposition where the goal is to keep the decomposition simple while still staying close to the original unitary map. The Solovay-Kitaev algorithm \citep{SK-orig,dawson2005solovaykitaev} performs an approximate decomposition, but uses only a finite set of gates. Since continuously parametrized gates are not included, a long sequence of elementary gates are necessary to closely approximate an arbitrary unitary matrix. The authors are not aware of an efficient algorithm that performs approximate decomposition while keeping the resulting circuits shallow.

\subsection{Conclusion}
\label{sec:conclusion}

While the arbitrary structure of input datasets makes theoretical claims of quantum machine learning applications challenging, the widespread adoption of machine learning in virtually every business sector creates a strong research motivation to explore potential QML applications. While this incentive guides QML research to approach applications from the ``what can we do with the hardware we have?" perspective, this work investigates a framework which focuses on the functional question: ``with an ideal quantum computer, are there circuit ansätze in the given architecture which outperform classical comparisons with the same amount of resources?" The framework in this paper is not absolute, but provides flexibility to quickly analyze the space of possible ansätze for a particular number of qubits within a given QML architecture with considerably favorable scaling compared to other state-of-the-art software packages, and can be extended with different architectures, encoding protocols, and decoding protocols than the ones shown in this work. The goal of this work is to provide a practical tool to the QML research community that can assist in guiding researchers towards effective ansatz structures for particular applications, or prevent researchers from over-investing in QML designs which have no clear advantage. As a framework, this methodology would be for the following open questions:
\begin{itemize}
  \item For the datasets explored in this work, would there be some set of architecture decisions which would lead to an benefit? This would involve exploring different sized quantum circuit ansätze, different numbers of layers of quantum transformations, and different positions of the operations in the stack.
  \item Could better circuit ansätze be found for previous algorithms? For instance, the quantum kitchen sinks \citep{quantum_kitchen_sinks} algorithm is very amenable to this approach. In that work, classical data is mapped into a single layer of Rx gates, with quantum circuit ansätze preceding and/or following this parameterized layer; using this framework, one could test if there were more appropriate ansätze to improve results found in the original paper.
  \item In the event that a set of useful ansätze is found for a particular dataset and architecture, how do we approximate it using available hardware? While there is a rich body of research highlighted in this work concerning decomposition, at the time of writing there are only a few solutions for approximating decomposing quantum circuits while also taking into account the impact of noise~\citep{cincio-PRX2021} and this constitutes a promising field of research.
\end{itemize}

In classical machine learning, this work relates with Neural Architecture Search (NAS)~\citep{elsken-hutter-NAS-survey-JMLR2019} which aims to automatically find the best network model for a given task and dataset e.g., image classification~\citep{zoph-NAS-cifar10-ICLR2017}. Combination of neural and quantum architecture search would allow optimal design of hybrid classical-quantum architectures.

\section{Acknowledgments and Disclosure of Funding}
B{\'{a}}lint  M{\'{a}}t{\'{e}} was supported by the Swiss National Science Foundation under grant number FNS-193716 ``Robust Deep Density Models for High-Energy Particle Physics and Solar Flare Analysis (RODEM)". This work was also inspired through collaboration and connection formed through the Quantum Open Source Foundation (\url{https://qosf.org/}).

\appendix
\section{Quantitative results of Experiment \ref{exp:1}}
\begin{table}[H]
\begin{center}
\rowcolors{2}{gray!12}{white}
    \begin{tabular}{ccccc} \toprule
        {qubits} & {PennyLane} & {PyTorch} & {PyTorch} & {PyTorch} \\ 
                 &             &           & {$+$batching} &  {$+$batching, GPU} \\ \midrule
1 & 6.75e-02 $\pm$ 5.02e-03 &2.13e-02 $\pm$ 3.70e-03 & \textbf{7.08e-04}$\pm$ 8.17e-05 &1.94e-03 $\pm$ 8.01e-04 \\
2 & 2.04e-01 $\pm$ 8.44e-04 &2.78e-02 $\pm$ 7.95e-03 & \textbf{8.43e-04}$\pm$ 5.57e-05 &1.70e-03 $\pm$ 7.85e-05 \\
3 & 6.62e-01 $\pm$ 5.30e-03 &7.27e-02 $\pm$ 6.51e-02 & \textbf{8.88e-04}$\pm$ 2.35e-05 &1.78e-03 $\pm$ 6.43e-05 \\
4 & 2.81e+00 $\pm$ 4.48e-01 &2.97e-02 $\pm$ 5.41e-04 & \textbf{1.55e-03}$\pm$ 1.37e-03 &1.85e-03 $\pm$ 5.48e-05 \\
5 & 1.19e+01 $\pm$ 5.57e-01 &3.99e-02 $\pm$ 3.21e-04 & \textbf{1.60e-03}$\pm$ 1.58e-05 &1.95e-03 $\pm$ 1.15e-04 \\
6 & 5.28e+01 $\pm$ 1.55e+00 &6.62e-02 $\pm$ 7.45e-04 &2.79e-03 $\pm$ 4.59e-05 & \textbf{2.19e-03}$\pm$ 1.19e-04 \\
7 & 2.34e+02 $\pm$ 9.33e+00 &1.78e-01 $\pm$ 7.38e-03 &7.44e-03 $\pm$ 1.06e-04 & \textbf{2.63e-03}$\pm$ 9.95e-05 \\
8 & 1.10e+03 $\pm$ 5.05e+01 &9.62e-01 $\pm$ 7.60e-02 &3.45e-02 $\pm$ 3.97e-04 & \textbf{3.69e-03}$\pm$ 1.95e-04 \\
\midrule
9 &  &7.27e+00 $\pm$ 2.44e-01 &3.02e-01 $\pm$ 6.27e-02 & \textbf{9.69e-03}$\pm$ 1.58e-04 \\
10 &  &5.68e+01 $\pm$ 7.12e-01 &1.96e+00 $\pm$ 1.59e-01 & \textbf{5.75e-02}$\pm$ 7.01e-04 \\
        \bottomrule
   
    \end{tabular}
    \caption{Results of Experiment \ref{exp:1}. Time it takes for each training process to complete 1 epoch on a dataset of 32 points. Mean and standard deviation values are reported. The CPU used for the first 3 columns was a AMD Ryzen 9 5950X, while the GPU for last column was a Nvidia RTX 3090. All values have units of seconds. }
     \label{table:results}
         \end{center}
    \end{table}

\bibliography{bib, bib_mph} 
%%\nocite{*}

\end{document}